\documentclass[12pt]{article}

\usepackage{xurl}
\usepackage{iaw}
\usepackage{multirow}  

\usepackage[backend=biber,style=authoryear,maxcitenames=3,maxbibnames=99,uniquelist=false,uniquename=false,dashed=false,doi=false,isbn=false,url=false,eprint=false]{biblatex}
\nocaptionfileinfo  

\newcommand{\Var}{\mathrm{Var}}
\newcommand{\Cov}{\mathrm{Cov}}

\newif\ifincludeexhibits
\includeexhibitstrue  

\ifincludeexhibits\else
  \usepackage{xr-hyper}
\fi

\newcommand{\TabPostTwoThousandFourRows}{%
    Mean & 0.25 & 0.08 & 0.13 & 0.08 & 0.09 \\
    Median & 0.19 & 0.07 & 0.12 & 0.07 & 0.07 \\
    25th pct &  0.02 & -0.04 & -0.02 & -0.04 & -0.03 \\
    75th pct & 0.39 & 0.18 & 0.25 & 0.18 & 0.21 \\
    \addlinespace
    \% pos & \mc1r{80} & \mc1r{67} & \mc1r{71} & \mc1r{67} & \mc1r{68} \\
    Med.\ $t$ & 1.11 & 0.45 & 0.60 & 0.45 & 0.46 \\
    Med.\ SR & 0.26 & 0.11 & 0.14 & 0.11 & 0.11 \\
    \addlinespace
    Med.\ $\alpha$ & 0.23 & 0.09 & 0.17 & 0.09 & 0.13 \\
    Med.\ $t(\alpha)$ & 1.53 & 0.64 & 1.09 & 0.64 & 0.70 \\
    N & \mc1r{204} & \mc1r{170} & \mc1r{194} & \mc1r{170} & \mc1r{161} \\
}

\newcommand{\TabSurvivorRows}{%
    Cash-based oper.\ profitability     & 0.66 &  0.40 & 0.06 & 0.00 & \citeauthor{balletal2016} (\citeyear{balletal2016}) \\
    Oper.\ profit.\ (R\&D adj)          & 0.60 &  0.26 & 0.05 & 0.00 & \citeauthor{balletal2016} (\citeyear{balletal2016}) \\
    Realized-implied vol.\ spread       & 0.59 &  0.35 & 0.05 & 0.00 & \citeauthor{balihovakimian2009} (\citeyear{balihovakimian2009}) \\
    Off-season momentum                 & 0.54 &  0.13 & 0.05 & 0.00 & \citeauthor{hestonsadka2008} (\citeyear{hestonsadka2008}) \\
    Net external financing              & 0.48 &  0.15 & 0.04 & 0.00 & \citeauthor{bradshawrichardsonsloan2006} (\citeyear{bradshawrichardsonsloan2006}) \\
    Seasonal momentum (16yr+)           & 0.48 &  0.21 & 0.04 & 0.00 & \citeauthor{hestonsadka2008} (\citeyear{hestonsadka2008}) \\
    Gross profitability                 & 0.44 &  0.14 & 0.04 & 0.00 & \citeauthor{novymarx2013} (\citeyear{novymarx2013}) \\
    R\&D over market cap                & 0.41 &  0.33 & 0.03 & 0.00 & \citeauthor{chanlakonishoksougiannis2001} (\citeyear{chanlakonishoksougiannis2001}) \\
    Net equity financing                & 0.39 &  0.16 & 0.03 & 0.00 & \citeauthor{bradshawrichardsonsloan2006} (\citeyear{bradshawrichardsonsloan2006}) \\
    Operating leverage                  & 0.38 & -0.01 & 0.03 & 0.00 & \citeauthor{novymarx2011} (\citeyear{novymarx2011}) \\
    \mc1l{$\vdots$} & \mc1c{$\vdots$} & \mc1c{$\vdots$} & \mc1c{$\vdots$} & \mc1c{$\vdots$} & \\
    \midrule
    Mean, all 170 anomalies             & 0.08 & -0.04 & 0.01 & 0.00 & \\
    SD, all 170 anomalies               & 0.18 & 0.13 & 0.02 & 0.00 & \\
}

\newcommand{\TabGroupsRows}{%
    Momentum                  & 17 & 0.01 & 0.05 & -0.16 & 0.09 & 53 \\
    Profitability             & 9 & 0.25 & 0.24 & 0.09 & 0.44 & 78 \\
    Value / Fundamentals      & 15 & -0.02 & -0.04 & -0.11 & 0.10 & 47 \\
    Investment / Growth       & 36 & 0.09 & 0.06 & -0.02 & 0.18 & 69 \\
    Trading / Liquidity       & 13 & 0.09 & 0.11 & 0.03 & 0.13 & 85 \\
    Accruals / Accounting     & 18 & 0.09 & 0.06 & -0.02 & 0.19 & 72 \\
    Intangibles               & 12 & 0.04 & 0.00 & -0.02 & 0.07 & 50 \\
    Other                     & 50 & 0.10 & 0.11 & -0.03 & 0.18 & 72 \\
    \addlinespace \midrule
    All                       & 170 & 0.08 & 0.07 & -0.04 & 0.18 & 67 \\
}

\newcommand{\FigOneMeanTLongShort}{+0.48}
\newcommand{\FigOneMeanTLongMinusMarket}{-0.31}

\newcommand{\FigOneVarTLongShort}{1.09}
\newcommand{\FigOneVarTLongMinusMarket}{0.98}

\begin{document}

\title{What Useful Alphas?}

\author{Andrew Y. Chen\\Federal Reserve Board\thanks{Email: \texttt{andrew.y.chen@frb.gov}.  The views expressed herein are those of the authors and do not necessarily reflect the position of the Board of Governors of the Federal Reserve or the Federal Reserve System.} \and Ivo Welch\\UCLA\thanks{Email: \texttt{ivo.welch@anderson.ucla.edu}.}}

\date{Draft: \today}

\maketitle

\begin{abstract}
    This paper examines about 200 published long-short anomaly equity portfolios (\textcite{chenzimmermann2022}).  Over the period through 2005 (December 2005 and earlier) and across all stocks, their median zero-investment return was an impressive \textbf{48~bp} per month.  Using only post-2005 years (January 2006 onward) reduces this to \textbf{19~bp}.  Using only ``non-micro'' top-3,000 stocks in the top 90\% of market capitalization reduces this to \textbf{26~bp}.  Using only post-2005 \emph{and} non-micro stocks reduces this to \textbf{7~bp}.  Even modest allowances for luck or transaction costs would have eliminated even these 7~bp.  The evidence strongly suggests that published academic anomalies have been useless to non-micro-cap portfolio managers in the 21st century. Public stock markets were \emph{very} efficient.
\end{abstract}

\noindent\textit{Keywords:} Anomalies. Post-Publication Decay.

\noindent\textit{Target:} Financial Analysts Journal.

\clearpage

\section{Introduction}

From 1973 to 2016, top academic finance journals published more than 200 firm-level characteristics that have shown that they could predict the cross-section of stock returns.  These anomalies have ranged from momentum to value to accruals to asset growth to profitability.  The original papers reported impressive long-short returns --- often 50 to 100~bps per month --- with solid $t$-statistics.  Thus, an obvious question arises for equity portfolio managers: which of these anomalies has worked in the stocks that they could have traded in their portfolios in their lifetimes?

Our paper focuses on non-microcap managers\footnote{There is no unique definition of micro-cap.  We have seen categorizations of large-cap, mid-cap, and small-cap stocks ranging from the top 1,500 stocks to the top 4,000 stocks.  Our 3,000 stocks seem a good compromise.  We also considered and describe below variations.} and provides a simple answer.  In the approximately 200 anomaly portfolios available on the open-source \textcite{chenzimmermann2022} dataset, excluding stocks outside the top 90\% of market capitalization, and the period through 2005 (December 2005 and earlier), the median zero-investment return was not 48~bps per month, but only \textbf{7}~bps per month from January 2006 on.  (The median CAPM alpha was only \textbf{9}~bps per month, with a median $t$-statistic of \textbf{0.64}.)  And even minimal transaction costs would have eliminated even this.  With basic adjustments for random noise, we cannot reject the null hypothesis that all strategies were useless.

Our paper has two parts.  The first part discusses anomaly returns excluding micro-cap stocks after 2005.  Our primary domain is stocks within the top 90\% of total market capitalization, which in practice is close to the 3,000 largest stocks.  The returns were small and economically uninteresting, at around 7~bps per month.  A simple adjustment for random noise shrinks even the best survivors to nearly zero, and even minimal trading costs would erase what little remains.  The exceptions were a handful of profitability and financing signals that showed modestly higher returns.

The second part asks the natural follow-up question: if anomaly returns have been so small, why do we read so often about seemingly much more impressive returns?  A simple two-by-two decomposition gives one answer.  One dimension is the sample period (through 2005 vs.\ post-2005).  The other dimension is the stock universe (all stocks vs. top-90\% of market capitalization).  The cell that best approximates the original-paper setting --- through 2005, all stocks --- shows the aforementioned healthy median return of 48~bps per month.  The cell that matches the practitioner's reality in the 21st century --- i.e., post-2005, non-micro-cap stocks --- shows 7~bps.  Large mean returns required \textit{both} ``good old days'' data and the presence of microcap stocks.  Restricting the sample to post-2005 cuts the median return by about 60\%, and restricting the universe to the top 90\% of market capitalization cuts it by about one-half.  Together, the two restrictions reduce the median by about 85\% --- and no longer enough to survive even small transaction costs.

The remainder of this paper is organized as follows.  Section~2 describes the data and our universe filters.  Section~3 presents the main results for post-2005 returns excluding micro-cap stocks.  Section~4 provides the two-by-two decomposition by era and cap domain.  Section~5 reviews the related literature.  Section~6 concludes with numbered findings and practical implications.

\section{Data and Universe Filters}

Our anomaly data come from the open-source dataset of \textcite{chenzimmermann2022}, available at \url{openassetpricing.com}.  This dataset reproduces nearly all published cross-sectional return predictors using standardized code and methodology.  For each anomaly, the dataset provides monthly long-short portfolio returns formed by sorting stocks on the predictive characteristic following the procedure in the original papers, typically using quintile or decile sorts.

The original published papers typically use all CRSP-listed stocks --- including thousands of small, illiquid names --- and report results over each paper's specific sample period.  Most sample periods begin well before 2006, and many start before 1980.  This baseline, which we label ``Original Papers,'' represents the upper bound of reported anomaly performance.

We apply two types of filters, separately and jointly.  The first is a \textbf{rank filter} (\textbf{N}).  The second is a \textbf{percentage filter} (\textbf{\%}): we keep stocks that collectively account for the top X\% of total market capitalization, starting from the largest.  The rank filter imposes a hard cap on the number of stocks.  The percentage filter ensures that the retained universe captures most of the market's value.  Applied together, a stock must pass both screens.  In each month, we keep only the qualifying stocks and form our anomaly portfolios from this restricted universe.
\begin{itemize}
    \item \textbf{Standard (N3000, 90\%)}: the intersection --- top 3,000 stocks that also fall within the top 90\% of market cap.  This is our baseline ``large and liquid'' universe.  This is our main filter.
\end{itemize}
We do however also consider alternative filters:
\begin{itemize}
    \item \textbf{Rank Only (N3000)}: the top 3,000 stocks by market capitalization.  This excludes the smallest microcaps but retains a broad universe.
    \item \textbf{Pct Only (90\%)}: stocks comprising the top 90\% of total market capitalization.  This is a value-weighted screen that excludes the long tail of tiny firms.
    \item \textbf{Tight (N1000, 80\%)}: the top 1,000 stocks that fall within the top 80\% of market cap.  This approximates the Russell 1000 or a typical large-cap mandate.
\end{itemize}

Some signals are defined only on small subsets of the market. For example, IO\_Short\-Interest keeps only stocks with the top 1\% of short interest (\textcite{asquith2005short}). This sparsity can lead to poorly-behaved portfolios, once our liquidity filters are applied.

We handle sparse signals in two steps. First, we apply signal screens based on observability in data through 2005. For each filter, we drop the signal if more than 5\% of its available months in the 1986--2005 sample have fewer than 20 stocks in either the long or short leg.  (The denominator is the signal's own available months, so signals that begin after 1986 are not penalized for late starts.)

Second, in post-2005 data, we treat sparse signals as an implementation outcome. If either the long leg or the short leg contains fewer than 20 stocks in a given month, we use a long-short return of zero for the next month, effectively assuming that the investor declines to trade due to insufficient diversification.

These two treatments ensure that the post-2005 results are largely implementable in real time. Previous drafts used alternative treatments and found similar results.

We split the sample period into two eras:  \emph{through 2005} (December 2005 and earlier) and \emph{post-2005} (January 2006 onward).  This split captures the idea that the nature of financial markets changed with the rise of information technology (\parencite{chordiaetal2014}). The precise split date is unimportant, but it must be after 2001, when market prices began being quoted in decimals rather than fractions. We follow \textcite{chenvelikov2023}, who split at the end of 2005, with the intention of capturing algorithmic and high-frequency trading.

\section{Post-2005 Returns in the Investable Universe}
\label{sec:main}

This section presents our paper's only point. It describes the distribution of long-short returns across all anomalies, restricting attention to the post-2005 period under our stock domain filters.

\instbl{tbl:post2004}

Table~\ref{tbl:post2004} reports summary statistics.  In the original all-stock universe, the median anomaly still earned about 19~bps per month after 2005.  Applying our Standard filter (N3000, 90\%) reduced the median to 7~bps.  The Rank Only (N3000) filter, which does not impose a market-cap percentage screen, left a median of 12~bps.  The Tight filter (N1000, 80\%) reduced the median further to 7~bps.  The share of anomalies with positive long-short returns fell from roughly 80\% in the all-stock universe to 67\% in the Standard universe and 68\% in the Tight universe.

\insfig{fig:tstat.hist.post2004}

The interquartile range under the Standard filter ran from about $-4$ to $+18$~bps per month.  The median $t$-statistic across anomalies was only 0.45, and the median annualized Sharpe ratio was 0.11.  The median CAPM alpha was 9~bps per month, with a median alpha $t$-statistic of only 0.64.  By any conventional standard, the vast majority of anomalies were statistically indistinguishable from zero.  Figure~\ref{fig:tstat.hist.post2004} shows the full post-2005 distribution of signal-level $t$-statistics and compares it to the standard normal null.

The cross-section of $t$-statistics also speaks to how much of the remaining performance was luck rather than signal.  The diagnostic is not any single anomaly's $t$-statistic but the spread of $t$-statistics across all of them.  Each $t$-statistic is true signal plus sampling noise, and the noise has variance 1 by construction.  Thus, even if every anomaly's true return were zero, the $t$-statistics of $\sim$200 anomalies would still scatter with a cross-sectional variance of about 1 --- the dispersion of 200 dart-throwing monkeys.  Only spread beyond 1 is evidence of genuine differences across anomalies.  \textcite{efron2010} (see also \textcite{chendim2023}) turns this observation into a simple shrinkage adjustment, $r_{adj} = [\,1 - 1/\Var(\,t\,)\,]{\cdot}\, r$, in which the factor $1 - 1/\Var(\,t\,)$ is the share of the observed spread that is signal rather than luck (Appendix~\ref{app:shrinkage}).  Post-2005 in the Standard universe, the $\Var(\,t\,)$ of the long-short returns was 1.09 --- almost exactly what luck alone would have produced.  The signal share was therefore only $1 - 1/1.09 \approx 0.08$, and the luck-adjusted return of every anomaly collapsed to nearly zero.  Even the best survivor in Table~\ref{tbl:survivors}, with a raw return of 66~bps per month, had a luck-adjusted return of only 6~bps.

The numbers in Table~\ref{tbl:post2004} are economically small.  A median return of 7~bps per month is about 1\% per year \emph{before} trading costs, and even minimal trading costs would eliminate such returns.

\instbl{tbl:survivors}

Table~\ref{tbl:survivors} lists the anomalies with the highest average long-short returns in the Standard (N3000, 90\%) universe after 2005.  The list is dominated by profitability and financing signals --- cash-based operating profitability, operating profitability adjusted for R\&D, gross profitability, and net external and net equity financing.  The third-best performer, the realized-implied volatility spread, is the one volatility signal on the list.  Only two momentum variants remain:  off-season momentum and long-horizon seasonal momentum.  Even these top performers mostly earned below 1\% per month before empirical Bayes shrinkage.  The shrinkage is motivated by the dispersion in Figure~\ref{fig:tstat.hist.post2004}:  the right tail contains genuine-looking survivors, but it is still selected from a noisy cross-section.

\instbl{tbl:groups}

Table~\ref{tbl:groups} groups anomalies by economic category and reports summary statistics within each group.  Profitability is the only category with a healthy median return.  Momentum, value, and intangibles-based anomalies have been essentially dead in the Standard (N3000, 90\%) universe after 2005.

The survival of two momentum-seasonality variants may not be surprising.  Momentum in its various forms has long been recognized as among the most robust patterns in the cross-section, and such persistence --- whether from delayed price adjustment, investor underreaction, or recurring seasonal demand --- plausibly survives even as other anomalies are arbitraged away.  A recent survey \parencite{welch2026} finds that academics have the most faith in momentum and profitability, though they consider neither to be compensation for risk.

Nevertheless, the surviving momentum variants' returns of about 50~bps per month in the top 90\% of stocks were below the 1.0 to 1.5\% per month reported (for non-seasonal momentum) in \textcite{jegadeeshtitman1993} for the full stock universe.  Momentum also carries well-known crash risk \parencite{danielmoskowitz2016}, which is not reflected in average returns.

\section{The Anatomy of Published Alphas}
\label{sec:2x2}

The contrast between the all-stock returns through 2005 and the post-2005 large-cap returns is stark.  In our all-stock replication through 2005, the median anomaly earned about 48~bps per month.  In the top 90\% of stocks after 2005, that figure fell to about 7~bps --- a decline of roughly 85\%.  Where did the returns go?  Or, more pointedly: how did the published literature produce such impressive numbers in the first place?

\insfig{fig:ret.by}

Figure~\ref{fig:ret.by} plots the cross-anomaly mean return and mean CAPM alpha against broader stock-universe cutoffs in the post-2005 sample.  Both rise materially as the cap filter is relaxed.  The all-stocks point sits well above the large-cap screens, reinforcing the paper's main message: much of the remaining anomaly performance lives outside the investable universe of a large-cap manager.

We decompose the gap using a simple two-by-two framework.  One dimension is the sample period:  the period through 2005 (as in the original papers) versus post-2005 only.  The other is the stock universe:  all stocks (as in the original papers) versus the Standard filter (top 90\% of market capitalization).  Figure~\ref{fig:2x2} reports the median long-short return and the share of anomalies with positive returns in each of the four cells.

\insfig{fig:2x2}

Start in the upper-left cell (through 2005, all stocks):  a median of \textbf{48~bps per month}, with 99\% of anomalies positive --- our closest analogue to the original-paper environment.  Moving right (restricting to the top 90\% of market cap) cut the median to about \textbf{26~bps}.  Moving down (restricting to post-2005) cut it to about \textbf{19~bps}.  Moving to the lower-right cell --- large stocks, post-2005 --- reduced the median to about \textbf{7~bps}, with 67\% of anomalies positive.  With the top 1,000 stocks and 80\%, the median fell to about \textbf{7~bps}.  The contrast between the upper-left cell (the published literature) and the lower-right (the practitioner's reality) is the central finding of this paper.

The two effects are roughly multiplicative:  each one reduces returns substantially, and together they reduce returns by about 85\%.  This means there is no single villain.  A researcher who uses the period through 2005 but restricts to large stocks will find modest but nonzero returns.  A researcher who uses all stocks but restricts to the post-2005 period will find a similar picture.  The returns disappear only when both constraints bind simultaneously --- which is precisely the situation facing a practitioner running a large-cap portfolio today.

The time dimension captures two related forces.  The first is post-publication decay:  once an anomaly is published, sophisticated investors trade on it, compressing returns.  \textcite{mcleanpontiff2016} document a roughly 50\% decline in returns post-publication across a broad sample of anomalies, and \textcite{chenzimmermann2020} confirm this finding in a larger dataset.  The second force is the broader revolution in trading technology.  Decimalization (2001), the rise of algorithmic trading, and the explosion of electronic market-making reduced trading costs and made it easier for arbitrageurs to act on published signals.  The combined effect of these two forces is that returns available after 2005 are much smaller than those available in the 1970s, 1980s, or 1990s, regardless of whether a specific anomaly had been published by that date.

The size dimension reflects a well-known but underappreciated fact:  most of the action in anomaly portfolios comes from small and microcap stocks.  These stocks have wider bid--ask spreads, lower institutional ownership, and less analyst coverage, all of which allow mispricings to persist.  When we exclude them, the anomaly returns shrink because the large-cap stocks that remain are more efficiently priced.  This does not mean the anomalies were ``fake'' in small stocks --- it means that the returns were concentrated in a segment of the market that most institutional investors cannot practically access at scale.

The two-by-two decomposition offers a simple framework for evaluating any published anomaly result.  Whenever a paper reports impressive long-short returns, the reader should ask two questions:  What stock universe was used?  And what sample period?  If the answer is ``all CRSP stocks'' and ``1965 to 2010,'' the reported returns may be largely irrelevant for a large-cap portfolio manager operating in 2025.  This is not a criticism of the original research --- the anomalies were genuine statistical findings.  It is a reminder that statistical significance and economic implementability are different things.

\section{Literature}

Our work builds on (and partly synthesizes) two closely related papers.\footnote{It is also heavily related to \textcite{mcleanpontiff2016}, who document that anomaly returns decay by roughly 50\% post-publication but remain positive on average.  \textcite{chen2021decay} shows that anomalies are statistically ``true discoveries'' --- they are not artifacts of data mining.  Our findings are consistent with both results.  The anomalies were real, and they were not fake.  They were, however, traded away.  The practical import of a ``true discovery'' that earns 5~bps per month in the investable universe is limited.}  \textcite{chenzimmermann2022} provide the open-source anomaly dataset and show that liquidity screens reduce in-sample returns by about 30\%.  They do not, however, examine size-filtered returns in the post-2005 subperiod.  \textcite{chenvelikov2023} show that the average anomaly's expected return is essentially zero after trading costs and post-publication decay.  Their approach uses trading cost adjustments rather than direct size filters, and their individual-anomaly analysis does not restrict the universe to large stocks.

Our own paper combines both size and time dimensions into a single memorable framework.  Of course, our paper can be critiqued on its edges --- what if we had just done X?  However, our paper is about the big picture.  It is not about whether this or that trick can resuscitate performance.  That said, at least one ``trick'' is important and interesting enough to deserve mention upfront and thus a proper caveat for our results. Our analysis uses the standard monthly-rebalanced long-short portfolios.  Dynamic factor timing \parencite{haddadetal2020}, volatility-managed portfolios \parencite{moreiramuir2017}, or machine-learning combinations of anomalies \parencite{gukellyxiu2020} seem to increase some returns again.\footnote{It is also true that dynamic strategies --- with their implicit options --- are more difficult to assess.  See \textcite{goetzmannetal2007}.}

Trading costs are not negligible even in the era of penny spreads and zero commissions.  \textcite{chenvelikov2023} show that anomaly portfolios overweight stocks with wider-than-average spreads --- about four times the median NYSE spread --- and turn over roughly 40\% of their two legs each month, so a half-spread of even 25~bps implies a round-trip cost near 20~bps per month.  Averaged across 204 anomalies, they find that the mean long-short return net of costs is about $-1$~bp per month post-2005 under the original implementations, rising to only about 4~bps under cost-minimizing execution.  Combining many anomalies does not rescue the enterprise:  gross combination returns of 250 to 380~bps per month through 2005 fall to 0 to 20~bps net afterward.  These figures span the full stock universe, where gross returns exceed those in our large-cap universe.  The small gross returns of Section~\ref{sec:main} are therefore upper bounds on what a large-cap manager could actually net.

\section{Conclusion}
\label{sec:conclusion}

This paper has made simple points about published stock return anomalies:

\begin{enumerate}

    \item \textbf{Post-2005, in the top 90\% of stocks by market capitalization, the median published anomaly earned about 7~bps per month (1\% per year) \emph{before} transaction costs.}  The median $t$-statistic was 0.45, the median CAPM alpha was 9~bps per month, and only about two-thirds of the anomalies retained positive long-short returns.  (Table~\ref{tbl:post2004}, Figure~\ref{fig:tstat.hist.post2004}.)

    \item \textbf{Published alphas required both data through 2005 and microcap stocks.}  A two-by-two decomposition shows that restricting the sample period to post-2005 reduced median returns by about 60\%, and restricting the universe to the top 90\% of market capitalization reduced median returns by about one-half.  Together, the reduction was about 85\%.  (Figure~\ref{fig:2x2}.)

    \item \textbf{The handful of survivors --- predominantly profitability based --- were almost entirely accounted for by luck.}  Cash-based operating profitability, operating profitability adjusted for R\&D, and the realized-implied volatility spread were the best performers, with raw returns of the best strategies of about 59 to 66~bps per month --- well below their published in-sample values.  The profitability category as a whole averaged about 25~bps per month.  A simple selection-bias adjustment implies that even the very strongest anomaly earned only 6~bps per month.  Even this required shorting --- the long-only return was at most zero net of selection bias.  (Tables~\ref{tbl:survivors} and \ref{tbl:groups}.)

\end{enumerate}

For a practitioner managing a stock portfolio without microcaps, the implication is sobering.  The published anomaly literature, taken at face value, does not offer a menu of profitable trading strategies.  The anomalies were real --- they were ``true discoveries'' in the statistical sense \parencite{chen2021decay} --- but they have been largely traded away in the segment of the market that matters most for institutional investors.

Of course, we do not claim that no anomaly-based strategy can ever be profitable.  We claim only that the standard published anomalies, implemented as described in their original papers, offered little in the investable universe after 2005.

This does not mean that quantitative strategies are futile.  Dynamic factor timing \parencite{haddadetal2020,moreiramuir2017}, machine-learning combinations \parencite{gukellyxiu2020,freybergeretal2020}, and strategies that exploit proprietary or alternative data go beyond what we study here.  Our results speak to the specific question of whether the 200-plus anomalies documented in the academic literature, implemented as described in their original papers, work in large stocks today.  The answer is: not much.

All of our data come from the open-source \textcite{chenzimmermann2022} dataset at \url{openassetpricing.com}.  The reader can replicate and extend our analysis.  We encourage practitioners and researchers to examine their own universe definitions, rebalancing frequencies, and cost assumptions.  The era of anomaly alphas in large-cap stocks may be over, but the era of transparent, reproducible asset pricing research is just beginning.

\clearpage
\phantomsection\addcontentsline{toc}{section}{References}
\section*{References}
\printbibliography[heading=none]

@article{asquith2005short,
  title={Short interest, institutional ownership, and stock returns},
  author={Asquith, Paul and Pathak, Parag A and Ritter, Jay R},
  journal={Journal of Financial Economics},
  volume={78},
  number={2},
  pages={243--276},
  year={2005},
  publisher={Elsevier}
}

@article{chen2021decay,
  author    = {Chen, Andrew Y.},
  title     = {The Decay of Return Predictability: How Does It Vary Across Characteristics?},
  year      = {2021},
  note      = {Working Paper},
}

@article{chenzimmermann2022,
  author    = {Chen, Andrew Y. and Zimmermann, Tom},
  title     = {Open Source Cross-Sectional Asset Pricing},
  journal   = {Critical Finance Review},
  year      = {2022},
  volume    = {11},
  pages     = {207--264},
}

@article{chenvelikov2023,
  author    = {Chen, Andrew Y. and Velikov, Mihail},
  title     = {Zeroing In on the Expected Returns of Anomalies},
  journal   = {Journal of Financial and Quantitative Analysis},
  year      = {2023},
  volume    = {58},
  number    = {3},
  pages     = {968--1004},
}

@misc{chendim2023,
  author        = {Chen, Andrew Y. and Dim, Chukwuma},
  title         = {High-Throughput Asset Pricing},
  year          = {2023},
  eprint        = {2311.10685},
  archiveprefix = {arXiv},
  primaryclass  = {q-fin.PM},
  note          = {Working Paper},
}

@article{chenzimmermann2020,
  author    = {Chen, Andrew Y. and Zimmermann, Tom},
  title     = {Publication Bias and the Cross-Section of Stock Returns},
  journal   = {Review of Asset Pricing Studies},
  year      = {2020},
  volume    = {10},
  number    = {2},
  pages     = {249--289},
}

@article{chordiaetal2014,
  author    = {Chordia, Tarun and Subrahmanyam, Avanidhar and Tong, Qing},
  title     = {Have Capital Market Anomalies Attenuated in the Recent Era of High Liquidity and Trading Activity?},
  journal   = {Journal of Accounting and Economics},
  year      = {2014},
  volume    = {58},
  number    = {1},
  pages     = {41--58},
}

@article{danielmoskowitz2016,
  author    = {Daniel, Kent and Moskowitz, Tobias J.},
  title     = {Momentum Crashes},
  journal   = {Journal of Financial Economics},
  year      = {2016},
  volume    = {122},
  number    = {2},
  pages     = {221--247},
}

@article{freybergeretal2020,
  author    = {Freyberger, Joachim and Neuhierl, Andreas and Weber, Michael},
  title     = {Dissecting Characteristics Nonparametrically},
  journal   = {Review of Financial Studies},
  year      = {2020},
  volume    = {33},
  number    = {5},
  pages     = {2326--2377},
}

@book{efron2010,
  author    = {Efron, Bradley},
  title     = {Large-Scale Inference: Empirical Bayes Methods for Estimation, Testing, and Prediction},
  publisher = {Cambridge University Press},
  year      = {2010},
  series    = {Institute of Mathematical Statistics Monographs},
  volume    = {1},
  address   = {Cambridge},
}

@article{gukellyxiu2020,
  author    = {Gu, Shihao and Kelly, Bryan and Xiu, Dacheng},
  title     = {Empirical Asset Pricing via Machine Learning},
  journal   = {Review of Financial Studies},
  year      = {2020},
  volume    = {33},
  number    = {5},
  pages     = {2223--2273},
}

@article{goetzmannetal2007,
  author    = {Goetzmann, William N. and Ingersoll, Jonathan E. and Spiegel, Matthew and Welch, Ivo},
  title     = {Portfolio Performance Manipulation and Manipulation-Proof Performance Measures},
  journal   = {Review of Financial Studies},
  year      = {2007},
  volume    = {20},
  number    = {5},
  pages     = {1503--1546},
}

@article{haddadetal2020,
  author    = {Haddad, Valentin and Kozak, Serhiy and Santosh, Shrihari},
  title     = {Factor Timing},
  journal   = {Review of Financial Studies},
  year      = {2020},
  volume    = {33},
  number    = {5},
  pages     = {1980--2018},
}

@article{jegadeeshtitman1993,
  author    = {Jegadeesh, Narasimhan and Titman, Sheridan},
  title     = {Returns to Buying Winners and Selling Losers: Implications for Stock Market Efficiency},
  journal   = {Journal of Finance},
  year      = {1993},
  volume    = {48},
  number    = {1},
  pages     = {65--91},
}

@article{mcleanpontiff2016,
  author    = {McLean, R. David and Pontiff, Jeffrey},
  title     = {Does Academic Research Destroy Stock Return Predictability?},
  journal   = {Journal of Finance},
  year      = {2016},
  volume    = {71},
  number    = {1},
  pages     = {5--32},
}

@article{moreiramuir2017,
  author    = {Moreira, Alan and Muir, Tyler},
  title     = {Volatility-Managed Portfolios},
  journal   = {Journal of Finance},
  year      = {2017},
  volume    = {72},
  number    = {4},
  pages     = {1611--1644},
}

@article{hestonsadka2008,
  author    = {Heston, Steven L. and Sadka, Ronnie},
  title     = {Seasonality in the Cross-Section of Stock Returns},
  journal   = {Journal of Financial Economics},
  year      = {2008},
  volume    = {87},
  number    = {2},
  pages     = {418--445},
}

@article{balletal2016,
  author    = {Ball, Ray and Gerakos, Joseph and Linnainmaa, Juhani T. and Nikolaev, Valeri},
  title     = {Accruals, Cash Flows, and Operating Profitability in the Cross Section of Stock Returns},
  journal   = {Journal of Financial Economics},
  year      = {2016},
  volume    = {121},
  number    = {1},
  pages     = {28--45},
}

@article{balihovakimian2009,
  author    = {Bali, Turan G. and Hovakimian, Armen},
  title     = {Volatility Spreads and Expected Stock Returns},
  journal   = {Management Science},
  year      = {2009},
  volume    = {55},
  number    = {11},
  pages     = {1797--1812},
}

@article{bradshawrichardsonsloan2006,
  author    = {Bradshaw, Mark T. and Richardson, Scott A. and Sloan, Richard G.},
  title     = {The Relation between Corporate Financing Activities, Analysts' Forecasts and Stock Returns},
  journal   = {Journal of Accounting and Economics},
  year      = {2006},
  volume    = {42},
  number    = {1--2},
  pages     = {53--85},
}

@article{novymarx2013,
  author    = {Novy-Marx, Robert},
  title     = {The Other Side of Value: The Gross Profitability Premium},
  journal   = {Journal of Financial Economics},
  year      = {2013},
  volume    = {108},
  number    = {1},
  pages     = {1--28},
  doi       = {10.1016/j.jfineco.2013.01.003},
}

@article{welch2026,
  author    = {Welch, Ivo},
  title     = {Assessing Factors and the {CAPM} in 2026},
  year      = {2026},
  note      = {Working Paper},
  doi       = {10.2139/ssrn.6325199},
}

@article{chanlakonishoksougiannis2001,
  author    = {Chan, Louis K. C. and Lakonishok, Josef and Sougiannis, Theodore},
  title     = {The Stock Market Valuation of Research and Development Expenditures},
  journal   = {Journal of Finance},
  year      = {2001},
  volume    = {56},
  number    = {6},
  pages     = {2431--2456},
}

@article{novymarx2011,
  author    = {Novy-Marx, Robert},
  title     = {Operating Leverage},
  journal   = {Review of Finance},
  year      = {2011},
  volume    = {15},
  number    = {1},
  pages     = {103--134},
}

\clearpage
\phantomsection\addcontentsline{toc}{section}{Tables and Figures}
\section*{Tables and Figures}

\setstretch{1.1}

\clearpage

\begin{table}
    \setstretch{1.1}

    \tblcaption{tbl:post2004}{Anomaly Long-Short Returns Post-2005, by Universe Filter}

    \begin{ctabular}{l F{2} F{2} F{2} F{2} F{2}}
        \toprule
        Statistic & \mc1c{No Constraint} & \mc1c{Standard} & \mc1c{Rank Only} & \mc1c{Pct Only} & \mc1c{Tight} \\
        \midrule
        \TabPostTwoThousandFourRows
        \bottomrule
    \end{ctabular}

    \explain{Summary statistics for the cross-sectional distribution of mean monthly long-short returns across $\sim$200 anomalies, restricted to the post-2005 period.  The number of anomalies varies by universe filter, as we drop a signal if more than 5\% of its available months through 2005 have fewer than 20 stocks in either leg. Returns and alphas are in percent per month.  ``No Constraint'' uses each anomaly's original all-stock portfolio from the Chen-Zimmermann dataset without any market-cap filter.  The other columns apply the indicated market-cap filter.  Med.~$t$ is the median $t$-statistic; Med.~SR is the median annualized Sharpe ratio; Med.~$\alpha$ and Med.~$t(\alpha)$ are the median CAPM alpha and its $t$-statistic.  Standard and Pct Only give identical results because the 90\% market-cap screen is more restrictive than the 3,000-stock rank screen.}

    \interpret{In the Standard (N3000, 90\%) universe, the median anomaly earned 7~bps per month --- about 1\% per year before trading costs.  About two-thirds of anomalies retained positive returns.}

\end{table}

\begin{figure}
    \setstretch{1.1}

    \figcaption{fig:tstat.hist.post2004}{Post-2005 N3000-90\% Signal-Level $t$-Statistics}

    \begin{ctabular}{cc}
        \textbf{Panel A: Long -- Short} & \textbf{Panel B: Long -- Market} \\
        \addlinespace
        \includegraphics[width=0.46\textwidth]{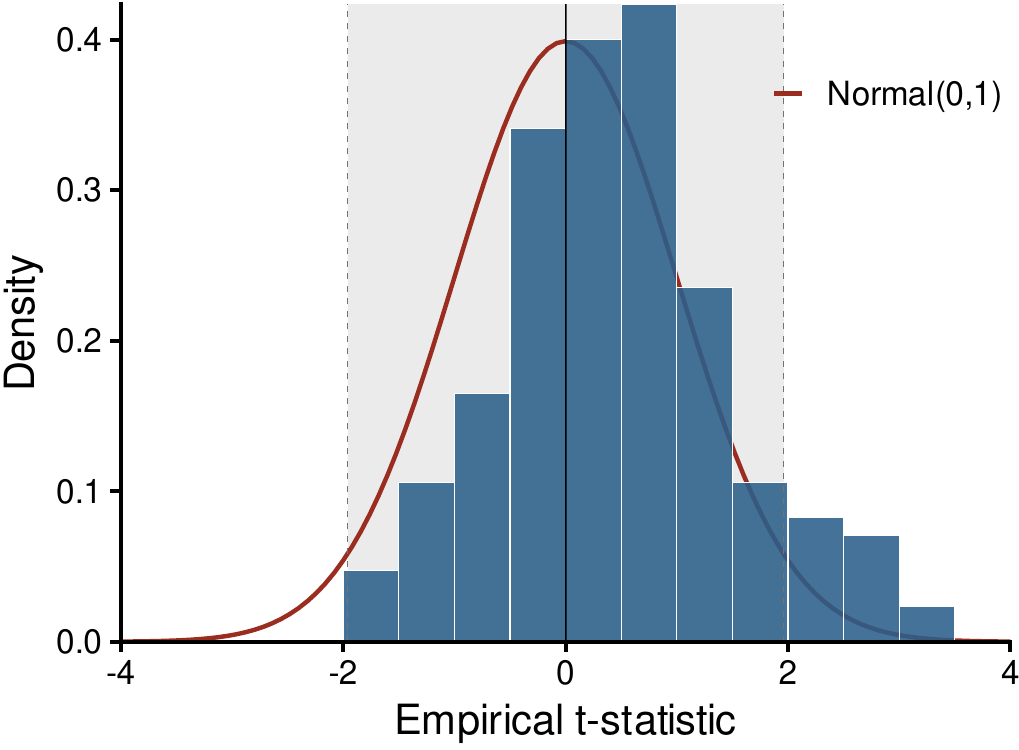} &
        \includegraphics[width=0.46\textwidth]{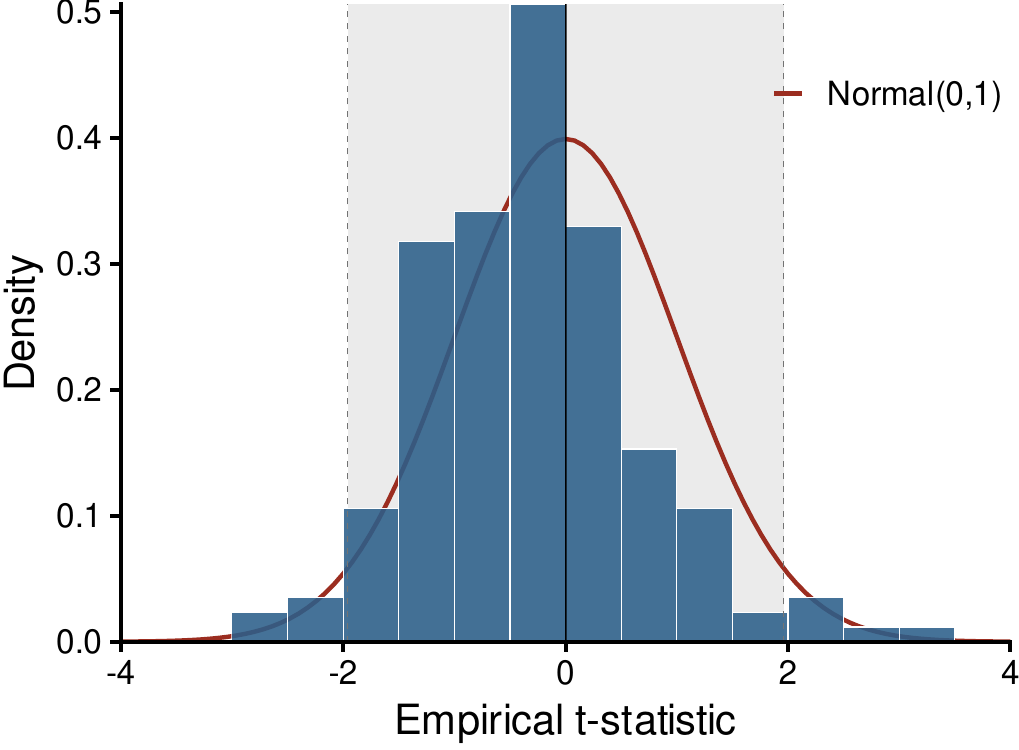} \\
        \addlinespace
        Mean $t$-statistic: $\FigOneMeanTLongShort$ & Mean $t$-statistic: $\FigOneMeanTLongMinusMarket$ \\
    \end{ctabular}

    \bigskip

    \explain{Each histogram shows post-2005 signal-level $t$-statistics in the Standard (N3000, 90\%) universe.  Long -- Short is the zero-investment anomaly return.  Long Minus Market is the return on each anomaly's long leg minus the raw Fama-French market return.  The red curve is the standard normal density.  The shaded region marks the conventional $[-1.96 , +1.96]$ interval.}

    \interpret{Post-2005, the right tail of long-short returns deviates slightly from the standard normal, implying some amount of true non-zero expected returns. But long-minus-market returns are very close to the standard normal, implying any long leg outperformance is entirely due to luck.}

\end{figure}

\clearpage

\begin{figure}

    \figcaption{fig:ret.by}{Average Mean Returns of Strategies by Exclusivity}

    \begin{ctabular}{c}
        \textbf{Panel A: As Function of $N$} \\
        \includegraphics[width=0.82\textwidth]{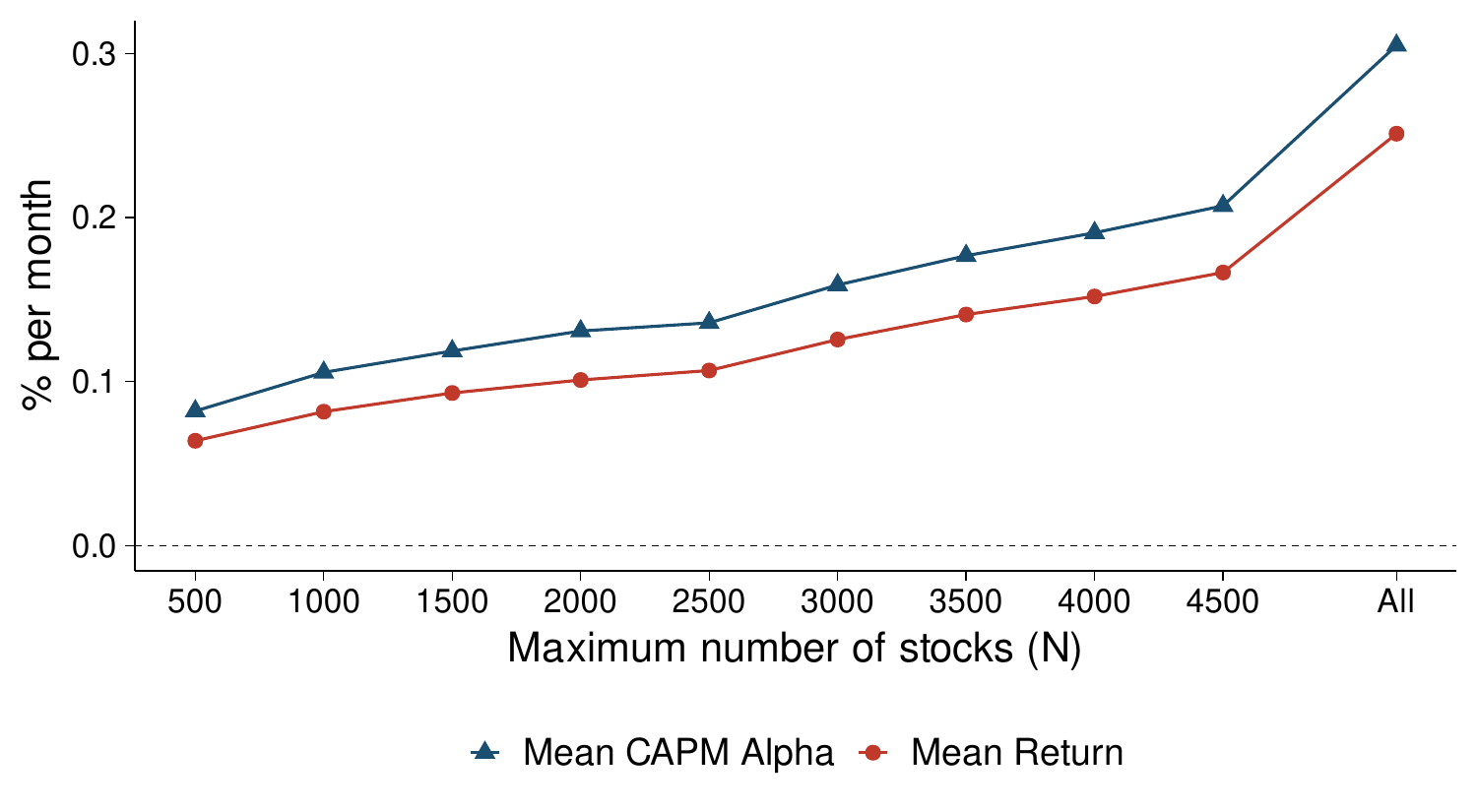} \\[1em]
        \addlinespace
        \textbf{Panel B: As Function of \% Excluded} \\
        \includegraphics[width=0.82\textwidth]{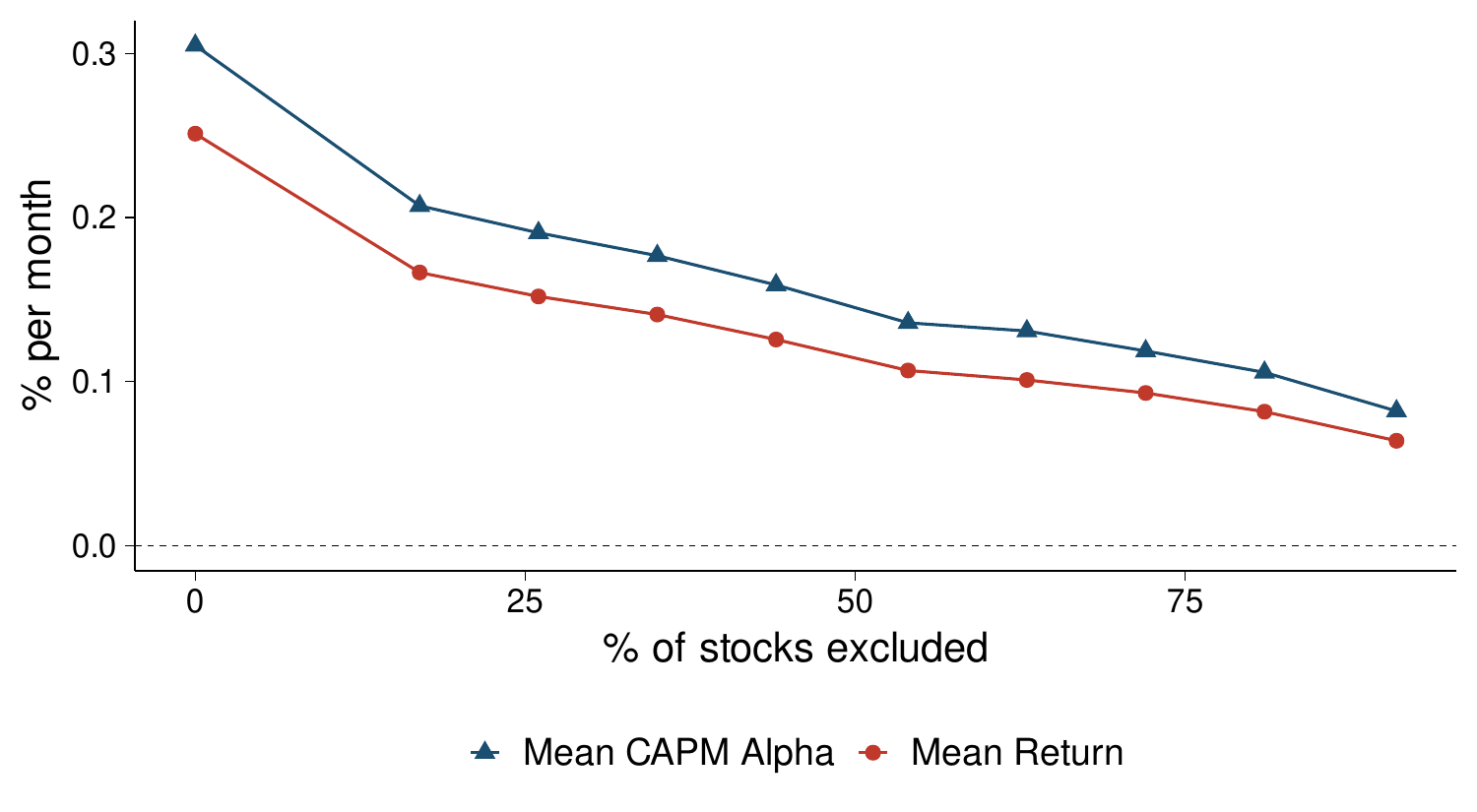} \\
    \end{ctabular}

    \explain{The x-axis is the number of stocks or percent of market cap included.  The y-axis is the mean zero-investment long-short return and the alpha of the 200 strategies.}

    \interpret{Predictability monotonically decreases in the liquidity of the set of stocks being considered.}

\end{figure}

\clearpage

\begin{table}
    \setstretch{1.2}

    \tblcaption{tbl:survivors}{Top 10 Anomalies Post-2005 in the Standard (N3000, 90\%) Universe}

    \begin{ctabular}{l F{2} F{2} F{2} F{2} >{\footnotesize} l}
        \toprule
        \multirow{2}{*}{Description} & \mc2c{Sample Mean} & \mc2c{Adjusted Mean} & \multirow{2}{*}{Original Paper} \\
        \cmidrule(lr){2-3}\cmidrule(lr){4-5}
        & \mc1c{LS $\downarrow$} & \mc1c{Long -- Mkt\hspace*{-1em}} & \mc1c{LS} & \mc1c{Long -- Mkt\hspace*{-1em}} & \\
        \midrule
        \TabSurvivorRows
        \bottomrule
    \end{ctabular}

    \bigskip

    \explain{Anomalies ranked by mean monthly long-short return in percent per month.  Mean Long -- Mkt is the mean signal-month return of the long leg minus the raw Fama-French market return, computed as Mkt -- RF plus RF.  Adjusted means are empirical Bayes estimates that shrink each raw mean toward zero, $r_{adj} = [\,1 - 1/\Var(\,t\,)\,]{\cdot}\, r$, with $\Var(\,t\,)$ estimated from the cross-section of post-2005 signal-level $t$-statistics, separately for LS and Long -- Mkt returns, and the shrinkage factor truncated at zero; see Appendix~\ref{app:shrinkage}.  For LS returns, $\Var(\,t\,) = \FigOneVarTLongShort$ implies a factor of about 0.084.  For Long -- Mkt returns, $\Var(\,t\,) = \FigOneVarTLongMinusMarket$ is below 1, so the factor is zero, and every adjusted mean is zero.  The last two rows report the cross-sectional mean and standard deviation of each column across all $\sim$170 anomalies, not only the top 10.  ``Original Paper'' is the primary reference for each signal.  Signals must have at least 30 post-2005 months.  IO\_ShortInterest is excluded from the ranking due to a suspected data construction issue.  MomRev is excluded as a curated overlap with the momentum survivor family.}

    \interpret{The survivors are predominantly profitability and financing signals.  Only two momentum-seasonality variants remain.  Even the best performers earn below 1\% per month before shrinkage, and their empirical Bayes adjusted means are at most 6~bps per month.}

\end{table}

\clearpage

\begin{table}
    \setstretch{1.3}

    \tblcaption{tbl:groups}{Anomaly Returns by Category, Post-2005, Standard (N3000, 90\%) Universe}

    \begin{ctabular}{l r F{2} F{2} ss F{2} F{2} r}
        \toprule
        Anomaly Category & \mc1c{N} & \mc1c{Mean $\downarrow$} & \mc1{css}{Median} & \mc1c{25th} & \mc1c{75th} & \mc1c{\% pos} \\
        \midrule
        \TabGroupsRows
        \bottomrule
    \end{ctabular}

    \explain{Anomalies are grouped by economic category following the classification in \textcite{chenzimmermann2022}.  Each row reports cross-sectional summary statistics of mean monthly long-short returns within that category, for the post-2005 period in the Standard (N3000, 90\%) universe.  Returns are in percent per month.  Category counts and boundaries are approximate.  Some anomalies span categories.}

    \interpret{Profitability is the only category with a healthy median return.  Momentum, value, and intangibles-based anomalies no longer outperformed in this universe.}

\end{table}

\clearpage

\begin{figure}
    \setstretch{1.1}

    \figcaption{fig:2x2}{The Two-by-Two, Visualized}

    \newcommand{\ig}[1]{\includegraphics[width=.48\linewidth]{#1}}

    \begin{ctabular}{lcc}
        & \panel{A}{Through 2005, All Stocks}          & \panel{B}{Through 2005, Top 90\% Mkt Cap}         \\
        & \ig{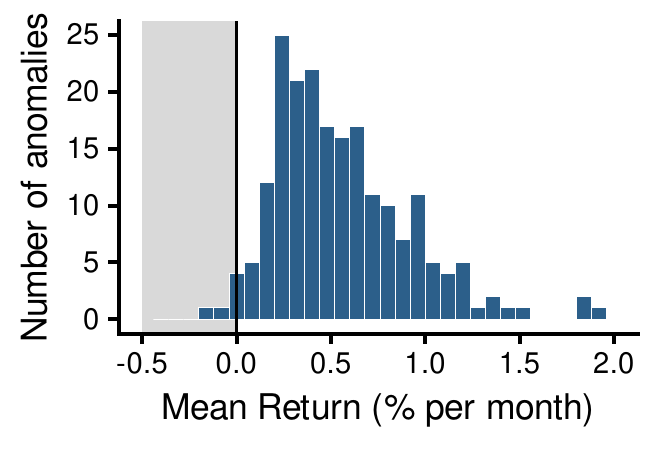} & \ig{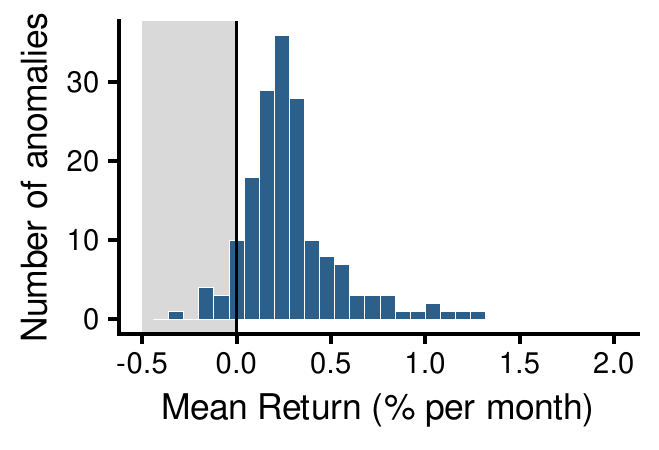} \\
        & Mean: 0.57\%                             & Mean: 0.28\%                                  \\
        & Median: 0.48\%                           & Median: 0.26\%                                \\
        & 99\% Positive                            & 92\% Positive                                 \\

        \addlinespace
        \addlinespace
        \addlinespace

        & \panel{C}{Post-2005, All Stocks}         & \panel{D}{Post-2005, Top 90\% Mkt Cap}        \\
        & \ig{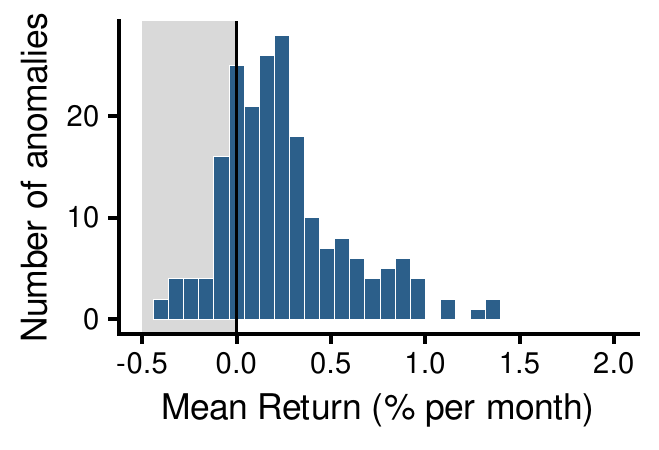} & \ig{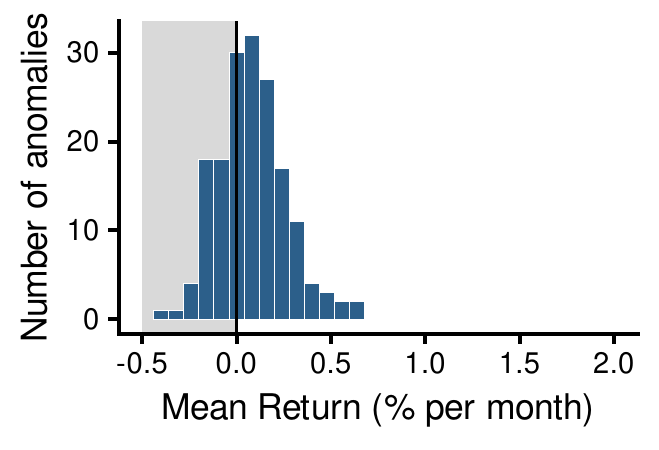} \\
        & Mean: 0.25\%                             & Mean: 0.08\%                                  \\
        & Median: 0.19\%                           & Median: 0.07\%                                \\
        & 80\% Positive                            & 67\% Positive                                 \\
    \end{ctabular}

    \explain{Each panel shows the distribution of mean monthly long-short returns across $\sim$200 anomalies.  The top row uses returns through December 2005.  The bottom row restricts to post-2005.  The left column includes all stocks.  The right column applies the Standard (N3000, 90\%) filter.}

    \interpret{Moving from the upper-left (published results) to the lower-right (practitioner reality), the distribution compresses dramatically toward zero.  The visual contrast is the central finding of this paper.}

\end{figure}

\clearpage
\appendix

\section{A Very Simple Bayes-Stein Shrinkage Formula}
\label{app:shrinkage}

To account for lucky performance among many predictors, a simple shrinkage adjustment is
\begin{equation}
    r_{adj} = \left[1 - \frac{1}{\Var(t)}\right] r .
    \label{eq:shrinkage-adjustment}
\end{equation}
Here $r$ is the mean return, $t$ is the $t$-statistic ($r$ divided by its standard error), and $\Var(t)$ is the cross-sectional variance of the $t$-statistics.

This expression is equivalent to Equation~(1.16) in \textcite[Chapter~1]{efron2010} and Equation~(8) of \textcite{chendim2023}.

\paragraph{Assumptions.}
Each predictor's $t$-statistic follows
\begin{align}
    t & = \theta + \delta ,
    \label{eq:shrinkage-tstat-model}
\end{align}
where $t$ is the observed $t$-statistic, $\delta$ is standard-normal noise, and $\theta$ is the latent true signal $t$-statistic.  We assume $\theta$ is normal with mean zero, and that $\theta$ and $\delta$ are independent within and across predictors.

\paragraph{Derivation.}
To remove the effect of noise $\delta$ conditional on a high $t$, compute the conditional expectation:
\begin{equation}
    E(t-\delta \mid t )
    = E(\theta \mid t)
    = \left[\frac{\Cov(\theta, t)}{\Var(t)}\right] t
    = \left[1 - \frac{1}{\Var(t)}\right] t ,
    \label{eq:shrinkage-conditional-expectation}
\end{equation}
where we use $\Var(t) = \Var(\theta) + \Var(\delta) = \Var(\theta)+1$ because $\theta$ and $\delta$ are independent.

Multiplying both sides by the standard error converts the result from $t$-statistics to returns.  If $\theta$ has a nonzero mean, the same logic shrinks each $t$-statistic toward the cross-sectional average rather than toward zero.  In our setting, the average $t$-statistic is small but positive (about 0.45); we return to the choice of shrinkage target below.

\paragraph{Illustration.}
To illustrate, consider three strategies with $t$-statistics of 0.5, 1.0, and 5.0.  Their cross-sectional variance is about 6.1.  Noise can account for only 1.0 of this 6.1, so about $5/6$ of the spread is signal.  The shrinkage factor is $1 - 1/6.1 \approx 0.84$:  an investor should believe most of what the data show and discount each strategy's mean return by only about 16\%.  Now consider instead three strategies with $t$-statistics of 0.5, 1.0, and 1.5 instead.  Their variance is 0.25 --- \emph{less} spread than luck alone would generate.  The formula then says the data contain no evidence of genuine differences across the strategies, and the best estimate of every strategy's deviation from the common mean is zero (the factor is truncated at zero).  Our $\sim$200 anomalies after 2005 were close to this second case:  $\Var(\,t\,) = 1.09$, so the shrinkage factor was 0.08.

\paragraph{Sign orientation.}
The \textcite{chenzimmermann2022} portfolios are signed so that each long-short return was positive in the direction the original paper documented.  This orientation does not affect $\Var(\,t\,)$ --- a variance is location-free --- but it does affect the shrinkage target.  With a nonzero cross-sectional mean, the empirical Bayes estimate shrinks each return toward the common mean rather than toward zero, $r_{adj} = \bar{r} + [\,1 - 1/\Var(\,t\,)\,]{\cdot}\,(\,r - \bar{r}\,)$.  Post-2005 in the Standard universe, the mean $t$-statistic was about 0.45, and the mean long-short return was 8~bps per month.  Shrinking toward this mean rather than toward zero would leave the median anomaly at about 8~bps per month and the best survivor at about 13~bps per month before trading costs.  The adjusted means in Table~\ref{tbl:survivors} instead shrink toward zero.  For the long legs net of the market, $\Var(\,t\,)$ was below 1, so the factor is truncated at zero either way, and the shrink-toward-mean estimate of every long leg equals the common average of about $-4$~bps per month.

It is not clear what one should shrink toward:  the cross-sectional mean or just zero.  The mean is not a neutral target.  These anomalies are a \emph{selected} set --- each was published because it worked, and each is signed in the direction in which it worked --- so their common mean is mechanically positive rather than a clean average of underlying premia.  Shrinking toward it imports that selection into the benchmark, whereas shrinking toward zero anchors on the null of no predictability.  The choice is not innocuous:  the best survivor's adjusted return is about 6~bps per month toward zero but about 13~bps toward the mean.

\end{document}